\documentclass{article}[12pt]
\usepackage{amssymb}
\usepackage{graphicx}
\usepackage{amsmath}

\newcommand{\be}{\begin{equation}}
\newcommand{\ee}{\end{equation}}
\newcommand{\bea}{\begin{eqnarray}}
\newcommand{\eea}{\end{eqnarray}}
\newcommand{\beas}{\begin{eqnarray*}}
\newcommand{\eeas}{\end{eqnarray*}}
\newcommand{\bi}{\begin{itemize}}
\newcommand{\ei}{\end{itemize}}
\newcommand{\bc}{\begin{center}}
\newcommand{\ec}{\end{center}}
\newcommand{\bfl}{\begin{flushleft}}
\newcommand{\efl}{\end{flushleft}}
\newcommand{\bfr}{\begin{flushright}}
\newcommand{\efr}{\end{flushright}}

\begin{document}

\title{Physical Variables of $d=3$ Maxwell-Chern-Simons Theory by Symplectic Projector Method}

\author{J. A. Helayel-Neto$^{a}$\footnote{E-mail: helayel@cbpf.br}, 
M. A. Santos$^{b}$\footnote{Email:masantos@ufrrj.br} and I. V. Vancea$^{c}$
\footnote{E-mail: ion@dfm.ffclrp.usp.br, ivancea@ift.unesp.br.}}

\date{\small $^a$ Centro Brasileiro de Pesquisas F\'{\i}sicas, (CBPF)\\R. Dr. Xavier Sigaud
150, 22290-180 Rio de Janeiro - RJ, Brasil \\
\vspace{.5cm}
$^b$Departamento de F\'{\i}sica, Universidade Federal do Espirito Santo (UFES)\\
Av. Fernando Ferrari S/N - Campos de Goiaberas, \\
29060-900 Vit\'{o}ria - ES, Brasil\\
\vspace{.5cm} 
$^c$
Departamento de F\'{\i}sica e Matem\'{a}tica,\\ Faculdade de Filosofia, Ci\^{e}ncias e Letras de Ribeir\~{a}o Preto
USP (FFCLRP-USP),\\ Av. Bandeirantes 3900, Ribeir\~{a}o Preto 14040-901, SP, Brasil
}

\maketitle

\abstract{
The Symplectic Projector Method is applied to derive the local 
physical degrees of freedom and the physical Hamiltonian of the Maxwell-Chern-Simons theory in $d=1+2$. 
The results agree with the ones obtained in the literature through different approaches.} 

\newpage

The quantization of the constrained systems is crucial in building realistic theoretical 
models. A major challenge in the process of quantization is the identification of the physical degrees
of freedom of the system. The most general method of quantization, the BRST method, reduces this problem
to the task of solving the cohomology of some nilpotent operators associated to the symmetry group \cite{BRST}.
Although elegant and complete, the full construction of BRST is sometimes unnecessary as more intuitive, albeit less
general methods, can be used for quantization. In \cite{Amaral1}, a procedure for separate the physical degrees of 
freedom for systems with second class constraints, called "symplectic projector method" (SPM), was proposed and 
it was subsequently developed in \cite{Amaral2, Marco1, Marco2, Marco3, Marcotese}. The idea behind the SPM is to 
construct a local projector from the phase space of the constrained system to the surface of constraints and to 
use it to obtain the local physical coordinates and the unconstrained Hamiltonian. The SPM represents a first 
step to treat the gauge theories in a strictly canonical way and it has already been applied to particles on 
holonomic surfaces \cite{Marco4}, non-comutative strings \cite{Marco5} and Abelian Chern-Simons systems \cite{Marco6}.

One of the most interesting class of models in field theory is described by the so called Maxwell-Chern-Simons theories
(MCS) which are important because they are simultaneously massive and gauge invariant. Recently, the MCS models have been used 
to study various phenomena related to the electric charges in the Standard 
Model Extension, topological massive electrodynamics and fractional statistics, vortex solutions in topological field 
theory, Lorentz symmetry breaking, $D$-brane Universe, large-$N$ field theories, dualities in field theories 
and quantum Hall effect to mention
just some of their applications. Therefore, the quantization of MCS theories represents an interesting
problem already addressed in the frameworks of the symplectic quantization \cite{Hong1}, geometric representation
\cite{Leal1}, covariant Coulomb gauge \cite{Haller1}, canonical Coulomb gauge \cite{Devecchi1}, Fadeev-Jackiew formalism
\cite{Lee1} and BFT formalism \cite{Fleck1} (see also \cite{Bekaert1, Ghosh1, Girotti1, Li1, Ghosh2, Itoh1}).

The aim of this letter is to explicitely derive the physical degrees of freedom and the physical Hamiltonian of the 
$d=3$ MCS theories by using the SPM in the canonical Coulomb gauge without matter. Our result 
is consistent with the one given in \cite{Devecchi1} which uses the Dirac quantization procedure. Compared to 
\cite{Devecchi1}, our approach is simpler and faster. This represents a non-trivial application of the SPM and proves 
that
it is an effective method applicable to interesting field theoretical models. 

Let us start by recalling the basic ideas of the SPM. Consider an arbitrary system with second class constraints
$\phi^m \left(\xi^M \right)=0$ where $\xi^M = (x^a, p_a )$, $M=1,2,\ldots,2N$ are the coordinates in the phase space 
which is assumed to be isomorphic to $R^{2N}$ and $m=1,2,\ldots, r=2k$. One can define a symplectic projector from the
phase space of the system to the constraint surface \cite{Amaral2} by the following relation
\begin{equation}
\Lambda ^{MN} = \;\; \delta ^{\,MN}- J^{ML}\,\frac{\delta{\phi}_{m}}{\delta
\xi ^{L}}\,\Delta^{-1}_{mn}\, \frac{\delta \phi _{n}} {\delta \xi^{N}}.
\label{m7}
\end{equation}
Here, $J^{MN}$ is the symplectic two-form in the original phase space and $\Delta^{-1}_{mn}$ is the inverse of 
the matrix constructed from the Poisson brackets of the constraint functions
\begin{equation}
\Delta_{mn} = \{\phi_m, \phi_n \}.  \label{submatrix}
\end{equation}
The action of the symplectic projector given by the relation (\ref{m7}) is to project the phase space variables 
$\xi^{M}$ onto a set of local variables on the constraint surface $\mathbf{\xi }^{*}$ 
\begin{equation}
\xi ^{*M}=\Lambda ^{MN}\xi ^{N}.  \label{m8}
\end{equation}
From these, one can construct the physical Hamiltonian by writing the original Hamiltonian in terms of the physical
coordinates (\ref{m8}) which are independent, unconstrained variables that obey the canonical commutation relations.
Next, one can derive the equations of motion from the Hamilton-Jacobi equations:
\begin{equation}
\stackrel{.}{\mathbf{\xi }}^{*}=\left\{ \mathbf{\xi }^{*},H^{*}\right\},
\label{m9}
\end{equation}
where $\{~,~ \}$ are the Poisson brackets. Comparing the Dirac matrix given by the following relation
\begin{equation}
D^{MN}=\{\xi ^{M}\;,\;\xi ^{N}\}_{D}=J^{MN}-J^{ML}J^{KN}\,\frac{\delta \phi
_{m}}{\delta \xi ^{L}}\,\Delta _{mn}^{-1}\,\frac{\delta \phi _{n}}{\delta
\xi ^{K}},  \label{m10}
\end{equation}
with the symplectic projector from (\ref{m7}), one can see \cite{Marco3} that the following relation holds: 
\begin{equation}
\Lambda =-DJ.  \label{m11}
\end{equation}
One can quantize the theory starting from the physical Hamiltonian described above. The other observables 
of the quantum theory are obtained in the same way and they depend on the physical coordinates only.

Now let us apply the above procedure to the MCS theory in Minkowski background. The Lagrangian is given by the 
following relation
\begin{equation}
\mathcal{L}=-\,\frac{1}{4}F_{\mu \,\nu }\,F^{\mu \,\nu }+m\varepsilon
^{\alpha \,\beta \,\gamma }\,A_{\alpha }\,\partial _{\beta }\,A_{\gamma }, 
\label{lagrangian}
\end{equation}
where $\mu, \nu = 1, 2, 3$, $\varepsilon$ is the antisymmetric tensor in $d=3$ and the metric has the
signature $\left( -1,1,1\right)$. Here, $F=dA$ and $m$ represents the mass parameter. The canonical Hamiltonian 
that is obtained from the Lagrangian (\ref{lagrangian}) has the following well known form
\begin{equation}
\mathcal{H}=\int d^{\,2}\,x\,\left[ \frac{1}{2}\pi \,^{i}\,\pi \,^{i}+\frac{1%
}{2}\left( \varepsilon ^{i\,j}\,\partial ^{\,i}\,A^{\,j}\right) ^{2}+\frac{1%
}{2}m^{2}\,A^{\,k}\,A^{\,k}+m\varepsilon \,^{i\,j}\,A^{\,i}\,\pi
^{\,j}\right].  \label{hamiltonian}
\end{equation}
The system displays second class constraints given by the following relations
\bea
\Omega ^{1}&=&\pi ^{\,0}=0, \label{constraint1}\\
\Omega ^{2}&=&\partial ^{\,i}\,\pi ^{\,i}+m\,\varepsilon ^{\,i\,j}\,\partial
^{\,j}\,A^{\,i}=0, \label{constraint2}\\
\Omega ^{3}&=&A^{0}=0,  \label{constraint3}\\
\Omega ^{4}&=&\partial ^{\,i}\,A^{\,i}=0. \label{constraint4}
\eea
The inverse $g_{ij}$ of the matrix 
\be
\ g^{i\,j}\,\left( x,y\right) =\left\{ \Omega ^{\,i}\,\left( x\right)
,\,\Omega ^{\,j}\,\left( y\right) \,\right\}, 
\label{inversemetric}
\ee
constructed from the above constraints (\ref{constraint1})-(\ref{constraint4}) defines a metric in the
phase space which has the following form
\begin{equation}
g^{-1}=\left( 
\begin{array}{cccc}
0 & 0 & \delta ^{\,2}\,\left( x-y\right) & 0 \\ 
0 & 0 & 0 & \nabla ^{-2} \\ 
-\,\delta ^{\,2}\,\left( x-y\right) & 0 & 0 & 0 \\ 
0 & -\,\nabla ^{-2} & 0 & 0
\end{array}
\right) .  \label{metric}
\end{equation}
By using the general formula (\ref{m7}), one can easily show that the local symplectic
projector in quantum field theory should be given by the following formula:
\begin{equation}
\Lambda _{\,\nu }^{\,\mu }\,\left( x,\,y\right) =\delta _{\,\nu }^{\,\mu
}\,\delta ^{\,2}\,\left( x-y\right) -\varepsilon ^{\,\mu \,\alpha }\,\int
d^{\,2}\,r\,d\,^{2}\,\varpi \,g_{i\,j}\,\left( r,\,\varpi \right) \,\delta
_{\,\alpha \,\left( \,x\right) }\,\Omega ^{\,i}\,\left( r\right) \,\delta
_{\,\nu \left( \,y\right) }\,\Omega ^{\,j}\,\left( \varpi \right) ,  \label{symproj}
\end{equation}
\medskip
where 
\be
\delta _{\,\alpha \,\left( \,x\right) }\,\Omega ^{\,i}\,\left(
r\right) \equiv \frac{\delta \,\Omega ^{\,i}\,\left( r\right) }{\delta \,\xi
^{\,\alpha }\,\left( x\right) }.\label{delta}
\ee

We now have all the ingredients at hand to find out the physical variables of the MCS theory
(\ref{hamiltonian}). As a first step we compute the symplectic projector of the system 
by explicitly working out (\ref{symproj}) and (\ref{metric}). After some tedious algebra one finds 
the following result
\be
\Lambda =\left( 
\begin{array}{cccccc}
0 & 0 & 0 & 0 & 0 & 0 \\ 
0 & \delta ^{2}\left( x-y\right) -\frac{\partial _{\,1}^{\,x}\,\partial
_{\,1}^{\,y}}{\nabla ^{\,2}} & -\,\frac{\partial _{1}^{x}\partial _{2}^{y}}{%
\nabla ^{2}} & 0 & 0 & 0 \\ 
0 & -\frac{\partial _{\,2}^{\,x}\,\partial _{\,1}^{\,y}}{\nabla ^{\,2}} & 
\delta ^{2}\left( x-y\right) -\frac{\partial _{\,2}^{\,x}\,\partial
_{\,2}^{\,y}}{\nabla ^{\,2}} & 0 & 0 & 0 \\ 
0 & 0 & 0 & 0 & 0 & 0 \\ 
0 & 0 & -\,m\,\delta ^{\,2}\left( x-y\right) & 0 & \delta ^{2}\left(
x-y\right) -\frac{\partial \,_{1}^{x}\,\partial _{1}^{y}}{\nabla ^{\,2}} & -%
\frac{\partial \,_{1}^{x}\,\partial \,_{2}^{y}}{\nabla ^{2}} \\ 
0 & m\,\delta ^{\,2}\left( x-y\right) & 0 & 0 & -\frac{\partial
\,_{2}^{x}\,\partial \,_{1}^{y}}{\nabla ^{2}} & \delta ^{2}\left( x-y\right)
-\frac{\partial \,_{2}^{x}\,\partial \,_{2}^{y}}{\nabla ^{2}}
\end{array}
\right) \,. \label{projectorMCS}
\ee
The next step is to apply the above projector to the field variables. Since the 
symplectic structure is most conveniently displayed in a symmetric notation, let us 
rename the field variables as follows
\begin{equation}
\left( A^{0},\,A^{1},\,A^{2},\,\pi ^{0},\,\pi ^{1},\,\pi ^{2}\right)
\Leftrightarrow \left( \xi ^{1},\,\xi ^{2},\,\xi ^{3},\,\xi ^{4},\,\xi
^{5},\,\xi ^{6}\right).  \label{newfields}
\end{equation}
The Hamiltonian (\ref{hamiltonian}) in this notation has the following form
\begin{equation}
\mathcal{H}=\int d^{\,2}\,x\,\left[ \,\frac{1}{2}\,\left( \xi \,_{5}^{2}+\xi
\,_{6}^{2}\right) +\frac{1}{2}\,\left( \partial _{1}\,\xi _{3}-\partial
_{2}\,\xi _{2}\right) ^{\,2}+\frac{1}{2}\,m^{2}\,\left( \xi \,_{2}^{2}+\xi
\,_{3}^{2}\right) +m\left( \xi \,_{2}\,\xi \,_{6}-\xi \,_{3}\,\xi
\,_{5}\right) \right].  \label{newhamiltonian}
\end{equation}
We denote the physical variables by $\xi _{\mu }^{\,\ast}\,\left( x\right) $. The definition (\ref{m8}) now
obviously reads
\begin{equation}
\xi ^{\,\mu \,\ast }\,\left( x\right) =\int d\,^{2}\,y\,\Lambda \,_{\nu
}^{\mu }\,\left( x,y\right) \,\xi ^{\nu }\,\left( y\right). \label{physicalfields}
\end{equation}
By using the relation (\ref{projectorMCS}) in to the equation (\ref{physicalfields}) one obtains the following
physical fields of the MCS theory
\bea
\xi ^{\,1\ast }\,\left( x\right) &=&\xi ^{\,4\ast }\,\left( x\right) =0, 
\label{physical1}\\
\xi ^{\,2\ast }\,\left( x\right) &=& A_{\,1}^{\,\perp }\,\left( x\right) , 
\label{physical2}\\
\xi ^{\,3\ast }\left( x\right) &=& A_{\,2}^{\,\perp }\,\left( x\right) , 
\label{physical3}\\
\xi ^{\,5\ast }\,\left( x\right) &=&\pi _{\,1}^{\,\perp }\,\left( x\right)
-m\,A_{\,2}^{\,\perp }\,\left( x\right) ,  
\label{physical4}\\
\xi ^{\,6\ast }\,\left( x\right) &=&\pi _{\,2}^{\,\perp }\,\left( x\right)
+m\,A_{\,1}^{\,\perp }\,\left( x\right) . 
\label{physical5}
\eea
By using the physical coordinates (\ref{physical1})-(\ref{physical5}) in the relation
(\ref{newhamiltonian}) we obtain the following projected Hamiltonian
\begin{equation}
\mathcal{H}^{\ast }=\int d^{2}x\left[ \frac{1}{2}\left( \xi _{5}^{\ast
2}+\xi _{6}^{\ast 2}\right) +\frac{1}{2}\left( \partial _{1}\xi _{3}^{\ast
}-\partial _{2}\xi _{2}^{\ast }\right) ^{2}+\frac{1}{2}m^{2}\left( \xi
_{2}^{\ast 2}+\xi _{3}^{\ast 2}\right) +m\left( \xi _{2}^{\ast }\xi
_{6}^{\ast }-\xi _{3}^{\ast }\xi _{5}^{\ast }\right) \right] .  
\label{projectedhamiltonian}
\end{equation}
From the projected Hamiltonian (\ref{projectedhamiltonian}) one derives the equations of motion by
using the Hamilton-Jacobi equations:
\bea
\stackrel{\ddot{}\ddot{}}{\xi }_{2}^{\ast } &=&-\,2\,m^{2}\,\xi \,_{2}^{\ast
}+\,\partial _{2}\,\partial _{2}\,\xi \,_{2}^{\ast }\,-\partial
_{1\,}\partial _{2}\,\xi \,_{3}^{\ast }-2\,m\,\xi \,_{6}^{\ast },  \label{em1}\\
\stackrel{\,\ddot{}\ddot{}}{\xi }_{\,3}^{\ast }&=&-\,2\,m^{2}\,\xi
\,_{3}^{\ast }+\partial _{1}\,\partial _{1}\,\xi \,_{3}^{\ast }-\partial
_{1}\,\partial _{2}\,\xi \,_{2}^{\ast }-2\,m\,\xi \,_{5}^{\ast },  \label{em2}\\
\stackrel{\,\ddot{}\ddot{}}{\xi }_{\,5}^{\ast }&=&\,-\,2\,m^{2}\,\xi
\,_{5}^{\ast }+\partial _{2\,}\partial _{2}\,\xi \,_{5}^{\ast }-\partial
_{1}\,\partial _{2}\,\xi \,_{6}^{\ast }+m\,\left[ \,2\,m^{2}-\nabla
^{2}\right] \,\xi _{3}^{\ast },  \label{em3}\\
\stackrel{\,\ddot{}\ddot{}}{\xi \,}_{6}^{\ast }&=&-\,2\,m^{2}\,\xi _{6}^{\ast
}+\partial _{1}\,\partial _{1}\,\xi _{6}^{\ast }-\partial _{1}\,\partial
_{2}\,\xi _{5}^{\ast }-m\,\left[ 2m^{2}-\nabla ^{2}\right] \,\xi _{2}^{\ast
}.  \label{em4}
\eea

In order to compare our results  with the ones given in the literature we go back to the standard
field notation in which the physical Hamiltonian (\ref{projectedhamiltonian}) has the form
\begin{equation}
\mathcal{H}^{\ast }=\int d^{2}\,x\,\left[ \frac{1}{2}\,\left( \pi
_{i}^{\perp }\,\pi _{i}^{\perp }+4\,m^{2}A\,_{i}^{\perp }\,A\,_{i}^{\perp
}\right) +\frac{1}{2}\left( \,\xi ^{\,i\,j}\,\partial _{\,i}\,A\,_{j}^{\perp
}\right) ^{2}+2\,m\,\left( A_{1}^{\perp }\,\pi \,_{2}^{\perp
}-A\,_{2}^{\perp }\,\pi _{1}^{\perp }\right) \right] .  \label{standardphysicalhamiltonian}
\end{equation}
\smallskip
This represents the MCS Hamiltonian in the canonical Coulomb gauge. The result
(\ref{standardphysicalhamiltonian}) agrees with the transverse expression of the
Hamiltonian obtained in \cite{Devecchi1} along a different line of arguments. Using the
standard field notations, the equations of motion given by the relations (\ref{em1})-(\ref{em4})
take the familiar look
\bea
\left( \square +4\,m^{\,2}\right) \,A_{\,1}^{\,\perp }=-\,2\,m\,\pi
_{\,2}^{\,\perp },  \label{sem1}\\
\left( \square +4\,m^{\,2}\right) \,A_{\,2}^{\,\perp }=\,2\,m\,\pi
_{\,1}^{\,\perp },  \label{sem2}\\
\square \,\pi _{\,1}^{\,\perp }=0,  \label{sem3}\\
\square \,\pi _{\,2}^{\,\perp }=0,  \label{sem4}
\eea
which amounts to ensuring that
\begin{equation}
\square \,\left( \square +4\,m^{2}\right) \,A_{\,i\,}^{\,\perp }=0,\,\,\ \ \
\left( i=1,2\right) .  \label{physicalequation}
\end{equation}
This equation  guarantees that the physical excitation is a massive $
\left( p^{2}=4\,m\,^{2}\right)$ transverse vector. The
massless quantum ($p^{2}=0$) is a spurious one: it has no 
dynamical r\^{o}le and does not correspond to any physical mode . Indeed, by
coupling the $A_{\mu }$ field propagator to a conserved
external current, the current-current amplitude is such that the imaginary
part of its residue taken at the pole $p^{2}=0$ vanishes, wich
confirms that the latter does not correspond to any physical
excitation. On the other hand, the non-trivial pole $p^{2}=4m^{2}$ yields a
positive defined residue which enforces its physical character as the only
degree of freedom carried by the $A_{\mu }$ field.

The quantization of the system should be performed in the usual fashion, starting 
from the physical
variables (\ref{physical1})-(\ref{physical5}). In order to avoid any confusion, we should stress out that 
the true physical Hamiltonian is the one given in the
relation (\ref{projectedhamiltonian}) in which the physical variables $\xi ^{\ast }$'s 
from (\ref{physical1})-(\ref{physical5}) obey the canonical Poisson brackets \cite{Amaral1}.
The transverse field variables used above just help us to compare the results 
obtained from the SPM with the ones in the literature. 

In conclusion, we have obtained the physical field variables and the physical Hamiltonian 
of the
MCS theory by projecting the constrained system onto the constraint surface. Our results 
agree with the
one obtained in the literature in the context of quantization of MCS within the canonical 
formalism. 

{\bf Acknowledgments}

We would like to thank to M. A. De Andrade for useful discussions. I. V. V. acknowledges discussions with
M. Botta-Cantcheff. I. V. V. was supported by FAPESP Grant 02/05327-3.

\end{document}